\documentclass{IEEEtran4PSCC}

\usepackage{ifpdf}

\usepackage{cite}

\ifCLASSINFOpdf
   \usepackage[pdftex]{graphicx}
\else
   \usepackage[dvips]{graphicx}
\fi
\usepackage[cmex10]{amsmath}
\usepackage{mathtools}
\usepackage{accents}
\newcommand{\ubar}[1]{\underaccent{\bar}{#1}}

\usepackage{algorithmic}

\usepackage{array}
\usepackage{fixltx2e}
 \usepackage{dblfloatfix}

 \usepackage{url}

\usepackage{longtable}
\usepackage[flushleft]{threeparttable}
\usepackage[flushleft]{threeparttablex}
\usepackage{multirow}
\usepackage{multicol}
\usepackage{subcaption}
\usepackage{xcolor}  
\usepackage{colortbl} 
\usepackage{booktabs} 
\usepackage{caption}  
\usepackage{acronym}
\usepackage{mathrsfs}
\usepackage{comment}

\usepackage{tikz}
\tikzset{
  panel label/.style={
    anchor=north west,
    font=\small,
    fill=white, fill opacity=0.75, text opacity=1,
    inner sep=2pt,
    xshift=3mm, yshift=-3mm
  }
}

\makeatletter
\g@addto@macro\normalsize{%
  \setlength{\abovedisplayskip}{0pt}
  \setlength{\belowdisplayskip}{0pt}
  \setlength{\abovedisplayshortskip}{0pt}
  \setlength{\belowdisplayshortskip}{0pt}
}

\makeatother

\acrodef{rvpp}[RVPP]{Renewable-based Virtual Power Plant}
\acrodef{vpp}[VPP]{Virtual Power Plant}
\acrodef{res}[RES]{Renewable Energy Sources}
\acrodef{ess}[ESS]{Electrical Storage System}
\acrodef{dam}[DAM]{Day-ahead Market}
\acrodef{asm}[ASM]{Ancillary Service Market}
\acrodef{idm}[IDM]{Intra-day Market}
\acrodef{srm}[SRM]{Secondary Reserve Market}
\acrodef{milp}[MILP]{Mixed Integer Linear Programming}
\acrodef{drs}[D-RES]{Dispatchable Renewable Energy Sources}
\acrodef{ndrs}[ND-RES]{Non-dispatchable RES}
\acrodef{csp}[CSP]{Concentrated Solar Power Plant}
\acrodef{sf}[SF]{Solar Field}
\acrodef{fd}[FD]{Flexible Demand}
\acrodef{wf}[WF]{Wind Farm}
\acrodef{pv}[PV]{Photo-Voltaic}
\acrodef{ts}[TS]{Thermal Storage}
\acrodef{mc}[MC]{Marginal Contributions}
\acrodef{sv}[SV]{Shapley Value}
\acrodef{der}[DER]{Distributed Energy Resources}
\acrodef{dso}[DSO]{Distribution Network Operator}
\acrodef{ro}[RO]{Robust Optimization}

\makeatletter
\let\old@ps@headings\ps@headings
\let\old@ps@IEEEtitlepagestyle\ps@IEEEtitlepagestyle
\def\psccfooter#1{%
    \def\ps@headings{%
        \old@ps@headings%
        \def\@oddfoot{\strut\hfill#1\hfill\strut}%
        \def\@evenfoot{\strut\hfill#1\hfill\strut}%
    }%
    \def\ps@IEEEtitlepagestyle{%
        \old@ps@IEEEtitlepagestyle%
        \def\@oddfoot{\strut\hfill#1\hfill\strut}%
        \def\@evenfoot{\strut\hfill#1\hfill\strut}%
    }%
    \ps@headings%
}
\makeatother

\begin{document}

%
\title{Enhancing Robust Multi-Market Participation of Renewable-Based VPPs through Flexible Resources\\}

\author{
\IEEEauthorblockN{Author n.1 Name per Affiliation A\\ Author n.2 Name per Affiliation A}
\IEEEauthorblockA{(Affiliation A) Department Name of Organization \\
Name of the organization, acronyms acceptable\\
City, Country\\
\{email author n.1, email author n.2\}@domain (if desired)}
\and
\IEEEauthorblockN{Author n.1 Name per Affiliation B\\ Author n.2 Name per Affiliation B}
\IEEEauthorblockA{(Affiliation B) Department Name of Organization \\
Name of the organization, acronyms acceptable\\
City, Country\\
\{email author n.1, email author n.2\}@domain (if desired)}
}

\author{
\IEEEauthorblockN{Hadi Nemati, {\'A}lvaro Ortega, Pedro S{\'a}nchez-Mart{\'i}n, Lukas Sigrist, Luis Rouco, Ignacio Egido}
\IEEEauthorblockA{Institute for Research in Technology, ICAI, Comillas Pontifical University, Madrid, Spain\\
\{hnemati, aortega, psanchez, lsigrist, rouco, egido\}@comillas.edu}
}



\maketitle

\begin{abstract}

In the transition toward a sustainable power system, \acp{rvpp} have emerged as a promising solution to the challenges of integrating renewable energy sources into electricity markets. Their viability, however, depends on effective market participation strategies and the ability to manage uncertainties while leveraging flexible resources. This paper analyzes the impact of different flexible resources—such as concentrated solar power plants, hydro plants, biomass plants, and flexible demand—on the participation of \acp{rvpp} in energy and reserve markets. Multiple sources of uncertainty in generation, consumption, and electricity prices are addressed using a two-stage robust optimization approach. The contribution of different technologies to \ac{rvpp} profitability is evaluated through a marginal contribution method, ensuring fair allocation of profits among them according to their actual role in energy and reserve provision across markets. Simulations for an \ac{rvpp} in southern Spain demonstrate how strategic decisions and the availability of flexible resources influence viability, market participation, and unit scheduling.

\end{abstract}


\thanksto{\noindent The authors wish to thank Comunidad de Madrid for the financial support to PREDFLEX project (TEC-2024/ECO-287), through the R\&D activity programme Tecnologías 2024.\vspace{-1.3em}}

\vspace{-1.2em}
\section{Introduction}

Coordinating diverse distributed resources has become essential for enhancing grid flexibility and enabling broader participation in electricity markets. In this context, the \ac{rvpp} emerges as a framework that unifies \ac{drs}, \ac{ndrs}, and \ac{fd}, leveraging digital infrastructure, advanced forecasting, and decentralized energy management~\cite{shafiekhani2022optimal}. {\color{black}By aggregating small-scale and distributed assets—often excluded from markets on their own—the \ac{rvpp} facilitates the integration of \acs{res} into wholesale markets, thereby potentially providing additional revenues~\cite{yang2023optimal}.} It also enhances operational flexibility for energy trading and reserve provision~\cite{mentens2025flexibility}. Recent reforms in Spain’s electricity market—such as integration with the European PICASSO platform~\cite{ENTSOE_PICASSO} and the expansion of demand response participation—have reshaped opportunities for \ac{rvpp} engagement~\cite{kaiss2025review}. These developments open new pathways for participation in the \ac{dam}, \ac{srm}, and \ac{idm}~\cite{marzbnai2025advances}. To exploit these opportunities, optimization models must represent the technical, operational, and economic characteristics of \acp{rvpp}, while also accounting for uncertainties in multi-market participation.

In contrast to traditional power systems—where thermal power plants dominate generation and fossil-fuel units provide the main source of flexibility—modern grids increasingly rely on \ac{ndrs}. {\color{black}Their weather-driven variability, which leads to forecast errors and limited dispatchability, introduces greater stochasticity compared to conventional portfolios~\cite{kaiss2025review}.} Achieving carbon-neutral targets therefore requires integrating clean flexibility on both supply and demand sides~\cite{mentens2025flexibility}. Recent studies highlight the role of flexible resources in improving \ac{rvpp} operation~\cite{shafiekhani2022optimal,yang2023optimal,li2024distributed,xiong2024distributionally}. Approaches include integrating distributed generation and dispatchable loads to enhance \ac{dam} participation of \ac{ndrs} under uncertainty~\cite{shafiekhani2022optimal}, using controllable sources such as \ac{ess} and hydropower to mitigate forecast errors~\cite{yang2023optimal, li2024distributed}, and combining \ac{csp} with \ac{ts} to improve dispatchability in wind-dominated \acsp{vpp}~\cite{xiong2024distributionally}.

The economic viability of flexible resources strongly depends on the \ac{rvpp}’s multi-market participation strategy and the role of each technology in shaping profitability. Ancillary service markets, such as the \ac{srm}, ensure reliability by securing sufficient reserve capacities in advance~\cite{van2020energy}. Several studies propose multi-market bidding and pricing strategies that integrate demand response, \ac{ess}, and electric vehicles to improve \acs{vpp} profitability and enhance the scheduling of \ac{ndrs}~\cite{feng2025optimal,chen2024pricing,nemati2025single}. Since day-ahead participation are increasingly affected by renewable uncertainty, \acp{idm} have emerged as complementary markets that allow continuous bid revisions based on updated forecasts~\cite{birkeland2024research}. Building on this link, recent works design scheduling, congestion management, and bidding mechanisms across the \ac{dam}, \ac{srm}, and \acp{idm}, leveraging demand flexibility, corrective actions, and rescheduling to reduce forecast errors, renewable curtailments, and exposure to market risks~\cite{nokandi2023three,kalantari2023strategic,wozabal2020optimal}.

Different methods have been proposed to allocate \acs{vpp} profits among members. Simple schemes, such as equal allocation and proportional sharing, divide profits uniformly or by unit size~\cite{gao2024review}, but fail to capture the real value of heterogeneous resources. To address this, contribution-based approaches such as the \ac{sv} and \ac{mc} have been developed. \ac{sv}-based methods allocate profits by quantifying each unit’s marginal impact on coalition performance and have been applied in energy, balancing, and regulation markets to better reflect unit contributions~\cite{dabbagh2015risk,zhang2023optimal}. The \ac{mc} method instead measures a unit’s importance by the profit reduction when it is excluded, offering computational simplicity and coalition stability. Recent works demonstrate its practicality by applying \ac{mc}-based allocation to distribution network operator–\acs{vpp} models and cost–benefit sharing under distribution constraints~\cite{huang2025co,yan2024multi}, making it especially appealing for real-world implementations where fairness and transparency are critical~\cite{gao2024review}.

{\color{black}Although several studies have examined different aspects of \acs{vpp} operation in electricity markets—such as uncertainty modeling and profit-sharing mechanisms—a comprehensive analysis of how various flexible resources affect \acs{rvpp} profitability across multiple markets (including the \ac{dam}, \ac{srm}, and \acp{idm}) under multiple sources of uncertainty is still lacking.} To address this gap, this paper examines an \ac{rvpp} composed of diverse flexible resources, including \ac{drs} (hydropower and biomass plants), \ac{csp} with \ac{ts}, and \ac{fd}, which together support the integration of \ac{ndrs} such as \ac{wf} and solar \ac{pv}. Uncertainty-handling strategies ranging from optimistic to pessimistic are analyzed within a two-stage \ac{ro} framework to demonstrate the impact of flexible resources on \ac{rvpp} profitability in multi-market participation. Furthermore, a \ac{mc}-based profit-allocation method is employed, as it balances fairness and practicality by reflecting the actual contribution of each technology to energy and reserve provision and to the overall viability of the \ac{rvpp}. Accordingly, the major contributions of this study include:

\begin{itemize}

\item {To evaluate the trading strategy and the energy and reserve scheduling of \ac{rvpp} units under different uncertainties—including renewable energy production, demand consumption, and multi-market electricity prices—and to assess the impact of a wide range of decision-making strategies, from optimistic to pessimistic, on trading and scheduling performance.}

\item {To analyze the multi-market participation of \ac{rvpp}, including the \ac{dam}, \ac{srm}, and \acp{idm}, and to evaluate the role of each market in the overall viability of the \ac{rvpp}.}

\item {To investigate the effect of various flexible resources and unit combinations—such as \ac{drs}, \ac{ndrs}, \ac{csp}, and \ac{fd}—along with different levels of demand flexibility on \ac{rvpp} profitability, {\color{black}and to develop a comprehensive profit-allocation strategy using \ac{mc} that accounts for the joint effects of market participation and uncertainty, thereby ensuring a fair distribution of profits according to the flexibility provided by each technology.}}

\end{itemize}

The paper is organized as follows: Section~\ref{sec:Problem_description} defines the the problem scope, while Section~\ref{sec:Formulation} presents the optimization framework for \ac{rvpp} multi-market participation. The validation of the model through case studies is presented in Section~\ref{sec:Case_studies}, and final insights are summarized in Section~\ref{sec:Conclusions}.

\section{Problem Description}
\label{sec:Problem_description}

\begin{figure}[!t]
    \centering
    \includegraphics[width=1\linewidth]{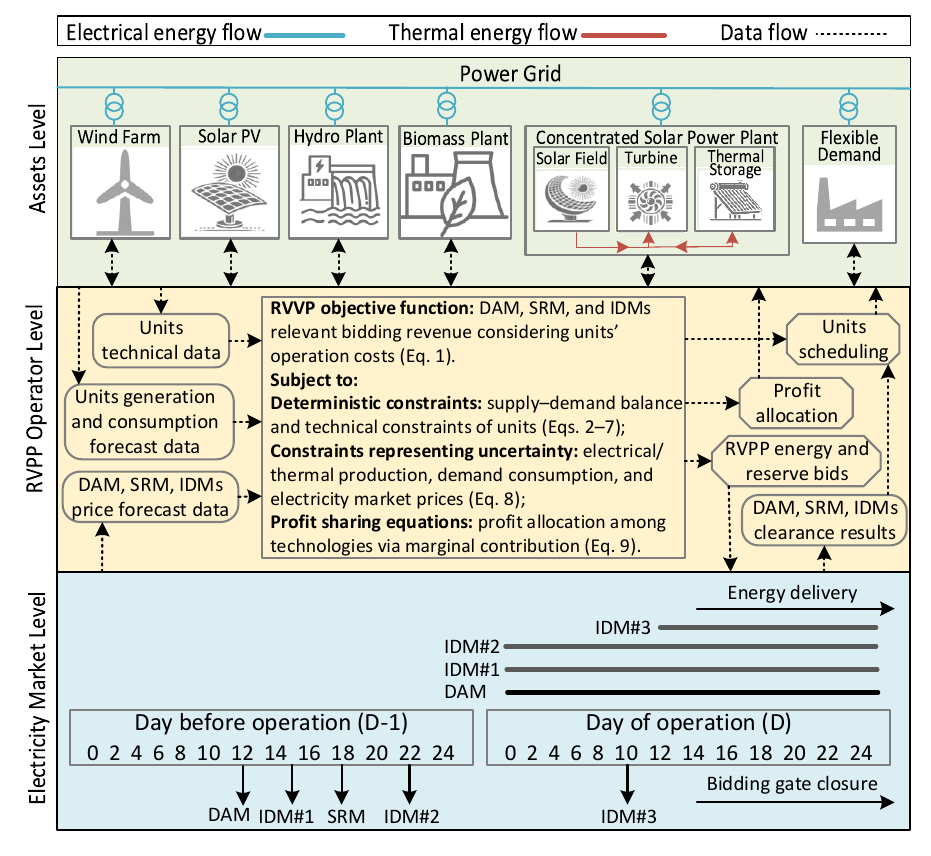}
    \vspace{-2em}
    \caption{The layout of the proposed \ac{rvpp}.}
     \label{fig:Scheme_RVPP}
     \vspace{-2em}
\end{figure}

Figure~\ref{fig:Scheme_RVPP} illustrates the schematic of the \ac{rvpp} participating in \ac{dam}, \ac{srm}, and \ac{idm}. The figure comprises three levels. At the asset level, multiple \ac{drs}, \ac{ndrs}, \ac{csp}, and \ac{fd} units are integrated into the \ac{rvpp} to enhance energy and service provision. At the \ac{rvpp} operator level, technical and forecast data are collected from units and markets to optimize scheduling and multi-market participation. The operator allocates profits among technologies according to their contribution in each market and the uncertainty-handling strategy adopted. The optimization framework provides bidding strategies for the relevant markets in which the \ac{rvpp} participates, as shown in the electricity market level. Depending on bidding gate closures, updated forecasts, and delivery times, the operator may co-optimize multiple products or fix the cleared results of previous markets in the optimization problem by adjusting the objective function. For instance, the operator can co-optimize energy and reserve before the \ac{dam} gate closure but only submit an energy bid to the \ac{dam}. Once the \ac{dam} is cleared, the operator can re-optimize participation in the \ac{srm} using updated forecasts and fixed energy bids from the \ac{dam}. The \ac{rvpp} is modeled as a price-taker, submitting zero-price bids to reflect its relatively small scale compared to the system. After receiving market-clearing results, the operator communicates dispatch instructions to its internal units.

{\color{black}The additional profit generated through aggregation, relative to independent operation, is allocated using the \ac{mc} method, enabling the contribution of each technology to be quantified by excluding it from the coalition. This paper focuses on technologies rather than individual units, aiming to identify which technologies—or their combinations—are most beneficial for the \ac{rvpp} operator. Unit-level allocation can also be performed if a more detailed comparison is required.} The bidding strategy of the \ac{rvpp} depends on its degree of conservatism toward uncertain parameters, which significantly affects profit allocation, particularly between dispatchable and non-dispatchable units. To address multiple uncertainties in multi-market participation under varying conservatism levels, this paper employs a two-stage \ac{ro} approach. This framework is suitable as it provides the operator flexibility to conduct multiple simulations in a computationally efficient manner.

\section{Formulation}
\label{sec:Formulation}
In this section, the deterministic model for \ac{rvpp} participation in the \ac{dam}, \ac{srm}, and \acp{idm} is first developed~\cite{nemati2025single}. Then, recognizing that price volatility, as well as generation and demand uncertainties, impact market outcomes, the model is extended to account for these uncertainties. Finally, the \ac{mc} method used for profit allocation between \ac{rvpp} units is explained. In~\eqref{RVPP: Obj_Deterministic}--\eqref{Deterministic: Reserve}, index $t \in \mathscr{T}$ denotes time periods; $r \in \mathscr{R}$ refers to \ac{ndrs}; $\theta\! \in\! \Theta$ to \acp{csp}; $c\! \in\! \mathscr{C}$ to \ac{drs}; $d \in \mathscr{D}$ to \acp{fd}; $m\! \in\! \mathscr{M}$ to load profiles; and $u\! \in\! \mathscr{U}$ to \ac{rvpp} units. Parameters $\lambda$, $P$, $R$, $C$, $T$, $K$, $\eta$, and $\Delta t$ represent market prices, electrical or thermal power capacity, reserve ramp rate, operating costs, reserve activation time, start-up multiplier, efficiency, and the duration of time periods, respectively. Decision variables $p$, $r$, and $e$ correspond to electrical or thermal power, reserve, and energy, respectively, while $u$ and $v$ are binary variables indicating the status and start-up of units. Superscripts and subscripts $E$, $R$, $\uparrow$, $\downarrow$, $SF$, $TS$, $\bar{A}$, and $\ubar{A}$ denote the energy market (\ac{dam} and \ac{idm}), reserve market (\ac{srm}), upward and downward regulation, the \ac{sf} of \ac{csp}, the \ac{ts} of \ac{csp}, and the upper and lower bounds of variables or parameters, respectively.

\subsection{Deterministic Problem}
\label{subsec:Deterministic_Formulation}

\subsubsection{Objective Function}
\label{subsubsec:Objective_Function}
The \ac{rvpp} objective function~\eqref{RVPP: Obj_Deterministic} maximizes total profit in the energy and reserve markets, while accounting for the operating expenses of its units. The terms in~\eqref{RVPP: Obj_Deterministic} represent, respectively: revenues from energy market bids; revenues from upward and downward \ac{srm} participation; and operating costs of \ac{ndrs}, \acp{csp}, and \ac{drs}.

\begingroup
\allowdisplaybreaks
\vspace{-.5em}
\begin{align} \label{RVPP: Obj_Deterministic}
&\mathop {\max }\limits_{{x \in X}} \sum\limits_{t \in \mathscr{T}} { {\lambda _t^{E}p_t^{E}\Delta t   } } + \sum\limits_{t \in \mathscr{T}} {\left[ {{\lambda _t^{{R, \uparrow}}r_t^{R,\uparrow} }  +{\lambda _t^{{R, \downarrow}}r_t^{R,\downarrow} }  } \right]}
 \\& - \sum\limits_{t \in \mathscr{T}} {\!\sum\limits_{r \in \mathscr{R}} {C_rp_{r,t}\Delta t} } \!- \sum\limits_{t \in \mathscr{T}} {\!\sum\limits_{\theta \in \Theta} {C_{\theta}p_{\theta,t}\Delta t} } \!- \sum\limits_{t \in \mathscr{T}} {\!\sum\limits_{c \in \mathscr{C}} { {{C_c}p_{c,t}\Delta t 
}  }}  \nonumber 
\end{align}
\vspace{.5em}
\endgroup

The objective function~\eqref{RVPP: Obj_Deterministic} can be adapted to each market session described in Section~\ref{sec:Problem_description}. When \ac{dam}+\ac{srm} is solved jointly, both energy revenues $(\lambda_t^{E} p_t^{E})$ and reserve revenues $(\lambda_t^{R,\uparrow} r_t^{\uparrow}$ and $\lambda_t^{R,\downarrow} r_t^{\downarrow})$ are active. In the \ac{srm} alone, only reserve revenues are considered, while energy revenues are fixed by \ac{dam} outcomes. For \ac{idm}\#1--\ac{idm}\#3, only energy trading terms apply and reserve terms are omitted. Cost terms are always included across all sessions.

\subsubsection{Supply--Demand Constraint}
\label{subsubsec:Supply_Demand_Constraints}

The supply--demand constraint of the \ac{rvpp} units is expressed in~\eqref{cons:Supply-Demand1}. {\color{black}This constraint contemplates, in a compact way, the expected real-time activation scenarios: upward, downward, and no activation~\cite{nemati2025single}. To capture this, reserve state vectors are introduced} as $\boldsymbol{r}_{t}^{R}=\{r_{t}^{R,\uparrow},-r_{t}^{R,\downarrow},0\}$ for the \ac{rvpp} and $\boldsymbol{r}_{u,t}=\{r_{u,t}^{\uparrow},-r_{u,t}^{\downarrow},0\}$ for each unit $u\in\mathscr{U}$, covering \ac{ndrs}, \acp{csp}, \ac{drs}, and \acp{fd}. As a result,~\eqref{cons:Supply-Demand1} expands into three distinct equations corresponding to the three activation scenarios.

\begingroup
\allowdisplaybreaks
\vspace{-.7em}
\begin{align}
    &\!\sum\limits_{r \in \mathscr{R}} \left[ p_{r,t} \!+ \boldsymbol{r}_{r,t} \right] \!+ \!\sum\limits_{\theta \in \Theta} \left[ p_{\theta,t} \!+ \boldsymbol{r}_{\theta,t} \right] \!+ \!\sum\limits_{c \in \mathscr{C}} \left[ p_{c,t} \!+ \boldsymbol{r}_{c,t} \right] 
    \nonumber \\& - \sum\limits_{d \in \mathscr{D}} \left[ p_{d,t} - \boldsymbol{r}_{d,t} \right] = p_{t}^{E}+ \boldsymbol{r}_{t}^{R}~;
    & 
    \forall t \label{cons:Supply-Demand1}
\end{align}
\vspace{.1em}
\endgroup

\subsubsection{Dispatchable Unit Constraints}
\label{subsubsec:TPP_Constraints}

Constraints~\eqref{Deterministic: DRES1} and~\eqref{Deterministic: DRES2} define the upper and lower bounds of \ac{drs} production based on the commitment binary variable $u_{c,t}$. 
The minimum up/down time ($UT_c$/$DT_c$) constraints are given in~\cite{carrion2006computationally} and are omitted here for brevity.


\begingroup
\allowdisplaybreaks
\begin{subequations}
\vspace{-.6em}
\begin{align}
    &p_{c,t} + r_{c,t}^{\uparrow} \leq \bar P_{c} u_{c,t}~; 
    &
    \forall c, t \label{Deterministic: DRES1} \\
    & \ubar P_{c} u_{c,t} \leq p_{c,t} - r_{c,t}^{\downarrow}~; 
    &
    \forall c, t \label{Deterministic: DRES2} 
    %
    %
    \end{align}    
\label{Deterministic: DRES}
\end{subequations}
\vspace{-.6em}
\endgroup

\subsubsection{Non-dispatchable Unit Constraints}
\label{subsubsec:RES_Constraints}

Constraint~\eqref{cons: NDRES1} defines the upper bound of \ac{ndrs} production using a fixed value for the uncertain parameter $P_{r,t}$. Constraint~\eqref{cons: NDRES2} specifies the lower bound on energy and reserve outputs~\cite{nemati2025single}.

\begingroup
\allowdisplaybreaks
\begin{subequations}
\vspace{-.6em}
\begin{align}
    & p_{r,t}+r_{r,t}^{\uparrow} \leq P_{r,t}~; 
    & 
    \forall r,t \label{cons: NDRES1} \\ 
    & \ubar P_{r} \le p_{r,t}-r_{r,t}^{\downarrow}~; 
    & 
    \forall r,t  \label{cons: NDRES2} 
%
%
    \end{align}
\label{RVPP: NDRES}
\end{subequations}
\vspace{-.8em}
\endgroup

\subsubsection{Concentrated Solar Power Plant Constraints}
\label{subsubsec:STU}

The \ac{csp} turbine converts thermal input from both the \ac{sf} and \ac{ts} into electricity as formulated in~\eqref{Deterministic: STU}~\cite{nemati2025two}. Constraint~\eqref{Deterministic: STU1} limits the thermal output of the \ac{sf}. Constraint~\eqref{Deterministic: STU2} ensures the balance between the turbine’s electrical energy output and its thermal energy input, accounting for \ac{sf} thermal power, \ac{ts} charging/discharging, and startup losses, while incorporating the turbine conversion efficiency ${\eta_{\theta}}$. Constraints~\eqref{Deterministic: STU3}--\eqref{Deterministic: STU4} restrict the \ac{csp}'s electrical output and reserves based on capacity limits and the binary commitment variable $u_{\theta,t}$. Constraint~\eqref{Deterministic: STU5} represents the thermal energy balance of the \ac{ts} over time, updating the energy level according to charging and discharging power together with their efficiencies. Minimum up/down time constraints for the \ac{csp} follow~\cite{carrion2006computationally} and are omitted for brevity.

\begingroup
\allowdisplaybreaks
\begin{subequations}
\vspace{-.7em}
\begin{align}
    & 0 \leq p_{\theta,t}^{SF} \leq {P}_{\theta,t}^{SF}~; 
    & \forall \theta, t \label{Deterministic: STU1} \\
    & \frac{p_{\theta,t}}{\eta_{\theta}} 
    = p_{\theta,t}^{SF} 
    + p_{\theta,t}^{TS, -} - p_{\theta,t}^{TS, +} 
    - K_{\theta} v_{\theta,t} \bar P_{\theta}~; 
    &\forall \theta, t \label{Deterministic: STU2} \\
    &p_{\theta,t} + r_{\theta,t}^{\uparrow} \leq \bar P_{\theta} u_{\theta,t}~; 
    &\forall \theta, t \label{Deterministic: STU3} \\
    &\ubar P_{\theta} u_{\theta,t} \leq p_{\theta,t} - r_{\theta,t}^{\downarrow}~; 
    &\forall \theta, t \label{Deterministic: STU4} \\
    & e_{\theta,t}^{TS} = e_{\theta,t-1}^{TS} + p_{\theta,t}^{TS,+} \eta_\theta^{TS,+} \Delta t - \frac{p_{\theta,t}^{TS,-} \Delta t}{\eta_{\theta}^{TS,-}}~; 
    &\forall \theta ,t \label{Deterministic: STU5}
    %
    %
    %
    \end{align} 
\label{Deterministic: STU}
\end{subequations}
\vspace{-.2em}
\endgroup

\subsubsection{Flexible Demand Constraints}
\label{subsubsec:Demand_Constraints}

The demand flexibility in this work enables the \ac{rvpp} to adapt its \ac{fd} behavior by either switching among profiles or adjusting within the limits associated with the chosen profile~\cite{ortega2022modeling}. Constraint~\eqref{cons: Demand1} establishes the minimum allowable load for \ac{fd}, accounting for profile-based variability. Constraint~\eqref{cons: Demand2} guarantees that only one profile is activated at a time. Meanwhile, the feasible domain for \ac{fd} operation—covering both consumption and reserve contributions—is assigned by~\eqref{cons: Demand3} and~\eqref{cons: Demand4}.

\begingroup
\vspace{-.7em}
\allowdisplaybreaks
\begin{subequations}
\begin{align}
    & p_{d,t} \geq \sum_{m \in  \mathscr{M}} \left[ {P}_{d,m,t} u_{d,m}  \right]~;
    & 
    \forall d,t \label{cons: Demand1}\\
    & \sum_{m \in \mathscr{M}} u_{d,m} = 1~; 
    & \forall d \label{cons: Demand2} \\
    & \ubar P_{d} \le p_{d,t} - r_{d,t}^{\uparrow}~;
    &
    \forall d,t  \label{cons: Demand3} \\
    & p_{d,t} + r_{d,t}^{\downarrow} \le \bar P_{d}~;
    & 
    \forall d,t  \label{cons: Demand4}
%
%
%
%
%
%
\end{align}
\label{RVPP: Demand}
\end{subequations}
\vspace{-1em}
\endgroup

\subsubsection{Reserve Provision Constraints}
\label{subsubsec:Reserve_Constraints}

Constraints~\eqref{Deterministic: Reserve1} and~\eqref{Deterministic: Reserve2} define the upward and downward reserves of \ac{rvpp} units based on their ramp-rate capabilities and the secondary reserve activation time.

\begingroup
\allowdisplaybreaks
\vspace{-1em}
\begin{subequations}
\begin{align}
    & r_{u,t}^{\uparrow} \le T^{R} R_{u}^{\uparrow}~;
    & 
    \forall u,t  \label{Deterministic: Reserve1} \\
    & r_{u,t}^{\downarrow} \le T^{R} R_{u}^{\downarrow}~;
    & 
    \forall u,t  \label{Deterministic: Reserve2}
    \end{align}    
\label{Deterministic: Reserve}
\vspace{-1.2em}
\end{subequations}
\endgroup

\subsection{Robust Problem}
\label{subsec:RVPP_Robust}

The robust \ac{rvpp} problem extends the deterministic formulation by accounting for uncertainties in market prices, renewable and solar-thermal generation, and demand. A two-stage \ac{ro} framework is adopted: in the first stage, the operator determines scheduling and market participation, while in the second, the worst-case realizations of uncertain parameters within prescribed sets are considered. The compact formulation is presented in~\eqref{Two-Stage_Robust}~\cite{nemati2025single}.

\begingroup
\allowdisplaybreaks
\begin{subequations}
\vspace{-1em}
\begin{align}
    & \max_{x \in X} \; \min_{\xi \in {\Xi}(\Gamma)} f(x,\xi)~; 
    & 
     \label{Robust_Obj} \\ 
    &  h(x) \leq 0~; 
    & 
     \label{Robust_Det} \\ 
    & g(x,\xi) \leq 0~; 
    & 
    \forall \xi \in {\Xi}(\Gamma)  \label{Robust_Unc}
    \end{align}
\label{Two-Stage_Robust}
\end{subequations}
\vspace{-.5em}
\endgroup

Where $x$ denotes first-stage decision variables, $\xi$ represents the uncertain parameters ${ \lambda_t^{E}, \lambda_t^{R,\uparrow}, \lambda_t^{R,\downarrow}, P_{r,t}, P_{\theta,t}^{SF}, P_{d,t} }$, and ${\Xi}(\Gamma)$ is the corresponding uncertainty set with budget $\Gamma$. The budget $\Gamma$ limits the number of deviations from nominal values, balancing robustness and conservatism: $\Gamma=0$ recovers the deterministic model, while the maximum $\Gamma$ provides full protection against all deviations. Each budget is an integer from 0 to 24 for each hour, allowing the conservatism level to range from optimistic to pessimistic. The function $f(x,\xi)$ in~\eqref{Robust_Obj} accounts for both first-stage decision variables in~\eqref{RVPP: Obj_Deterministic} and uncertain parameters related to electricity prices. The function $h(x)$ in~\eqref{Robust_Det} includes only deterministic decision variables and corresponds to the constraints in~\eqref{cons:Supply-Demand1}--\eqref{Deterministic: Reserve} that are unaffected by uncertainty. Finally, the function $g(x,\xi)$ in~\eqref{Robust_Unc} represents the constraints with uncertain parameters related to \ac{ndrs} electrical generation, \ac{csp} thermal production, and demand in~\eqref{cons: NDRES1},~\eqref{Deterministic: STU1}, and~\eqref{cons: Demand1}.

The two-stage problem~\eqref{Two-Stage_Robust} is reformulated as a single-level \ac{milp} using the standard strong duality principle~\cite{floudas1995nonlinear}, whose detailed formulation is omitted for brevity.

\vspace{-.25em}
\subsection{Profit Sharing via Marginal Contribution}
\label{subsec:profit_sharing}
The \ac{mc} method is adopted for profit allocation in the \ac{rvpp}~\cite{yan2024multi}. This approach is attractive because it reflects the actual role of each unit {\color{black}(or technology)} across markets: units that contribute more value to the coalition receive a larger share of the additional profit. Thus, the allocation balances fairness and practicality while considering both unit capacity and realized contribution. We denote by $\Pi^{\mathrm{RVPP}}$ the total \ac{rvpp} profit with all units included, by $\Pi_u^{\mathrm{solo}}$ the profit of unit $u$ when it participates individually in the market, and by $\Pi^{\mathrm{RVPP}\setminus u}$ the \ac{rvpp} profit when unit $u$ is excluded. Based on these definitions, the \ac{mc}-based allocation is given in~\eqref{Marginal_Contribution}.

 
\begingroup
\allowdisplaybreaks
\begin{subequations}
\begin{align}
    & \rho_u = \frac{\Pi^{\mathrm{RVPP}} - \Pi^{\mathrm{RVPP}\setminus u}}{P_u}; 
    && \forall u 
    \label{eq:rho} \\
    & \Delta \Pi = \Pi^{\mathrm{RVPP}} - \sum_{u \in \mathscr{U}} \Pi_u^{\mathrm{solo}}; 
    && 
    \label{eq:deltaPi} \\
    & \Pi_u^{\mathrm{alloc}} = \Pi_u^{\mathrm{solo}} 
    + \frac{\rho_u P_u}{\sum_{u \in \mathscr{U}} \rho_u P_u} \Delta \Pi; 
    && \forall u 
    \label{eq:alloc}
\end{align}
\label{Marginal_Contribution}
\end{subequations}
\endgroup

Equation~\eqref{eq:rho} defines the normalized marginal contribution of each unit ($\rho_u$), i.e., the reduction in overall \ac{rvpp} profit when that unit is removed, scaled by its capacity. Equation~\eqref{eq:deltaPi} captures the incremental ($\Delta \Pi$) profit that the \ac{rvpp} achieves relative to the sum of units acting individually. Finally, Equation~\eqref{eq:alloc} specifies the unit profit allocation ($\Pi_u^{\mathrm{alloc}}$): each unit receives its standalone profit plus a share of the incremental profit, weighted by $\rho_u P_u$. These weights ensure that larger units with higher marginal contributions receive proportionally greater allocations. Moreover, the proposed \ac{mc} method is budget-balanced, i.e., $\sum_{u \in \mathscr{U}} \Pi_u^{\mathrm{alloc}} = \Pi^{\mathrm{RVPP}}$.

\section{Case studies}\label{sec:Case_studies}

\begin{figure} [t!]
\vspace{-.1em}
    \centering 
    \includegraphics[width=\linewidth]{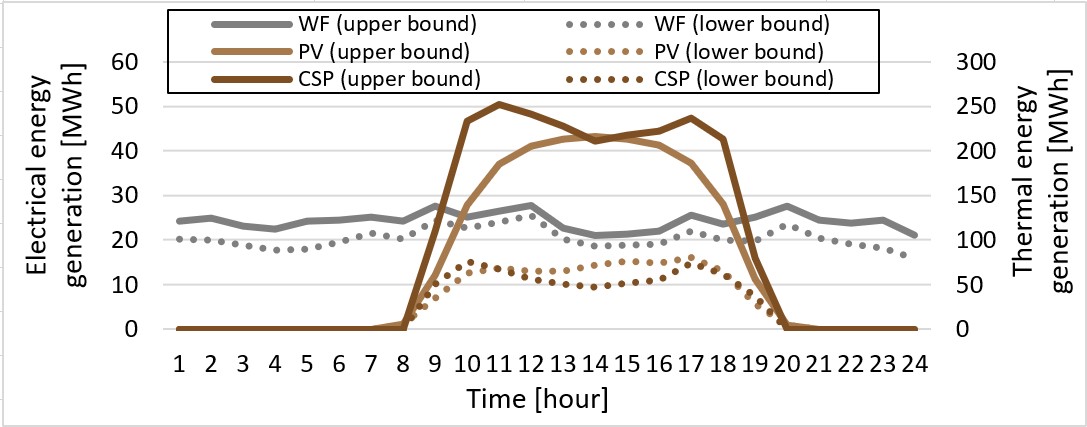}
    \includegraphics[width=\linewidth]{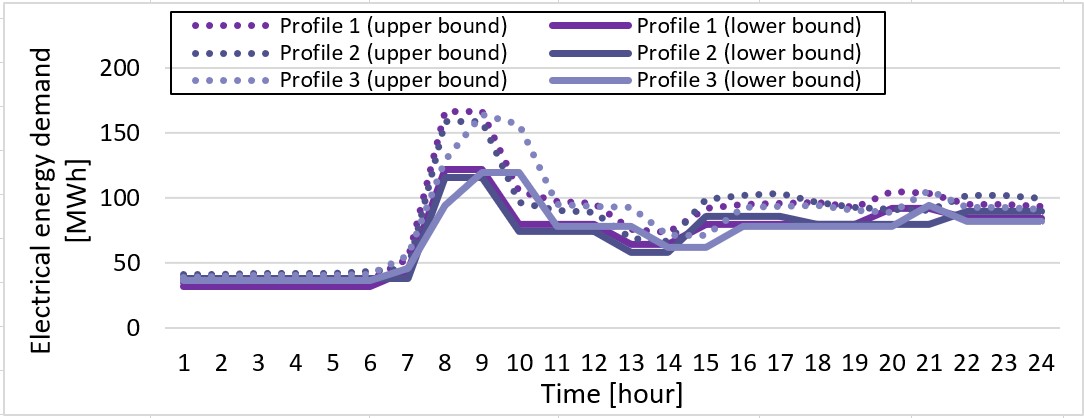}
    \caption{The forecast bounds of RVPP units.}
    \label{fig:Data_Production_Robust}
     \vspace{-.5em}
\end{figure}

\begin{figure} [t!]
\vspace{-.5em}
    \centering 
    \includegraphics[width=\linewidth]{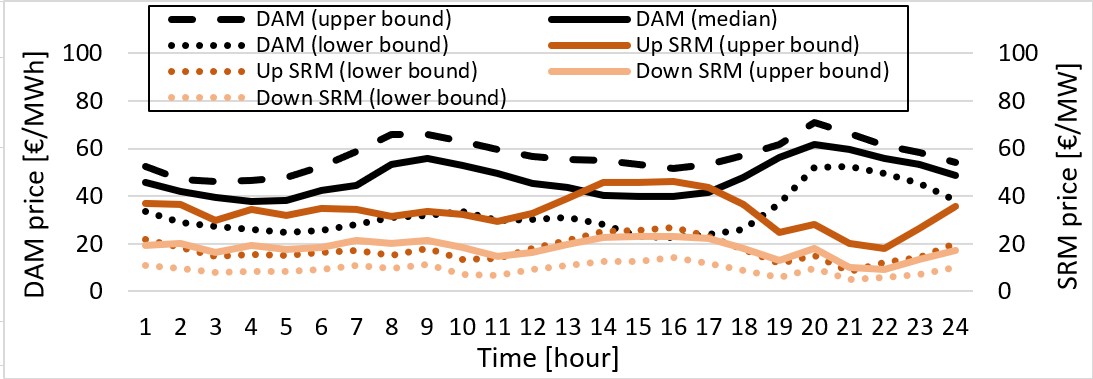}
    \includegraphics[width=\linewidth]{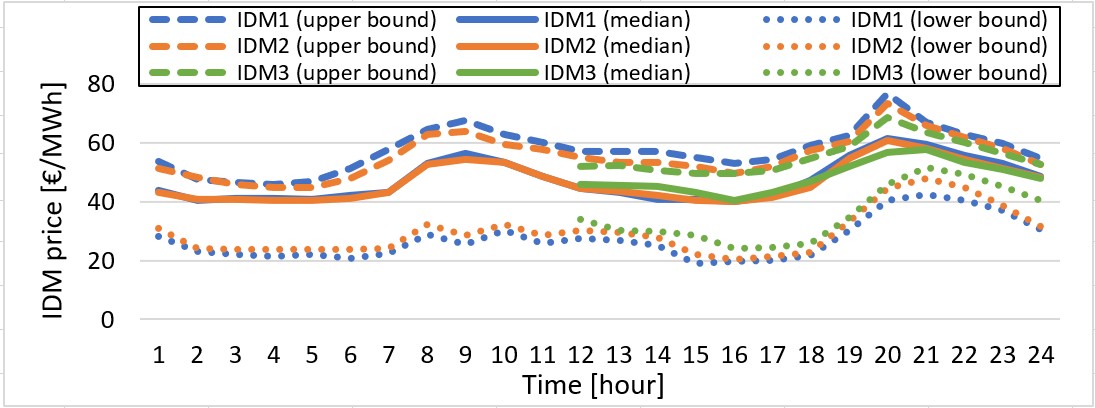}
    \caption{The forecast bounds of DAM, SRM, and IDMs price.}
    \label{fig:Data_Price_Robust}
    \vspace{-1em}
\end{figure}

This section presents simulation results based on the proposed framework to evaluate the impact of uncertainties and flexibility resources on \ac{rvpp} scheduling and profitability across electricity markets. Simulations are performed on a Dell XPS (i7-1165G7, 2.8 GHz, 16 GB RAM) using the CPLEX solver in GAMS 39.1.1. The case study considers an \ac{rvpp} in Southern Spain consisting of a hydro plant, a biomass unit, a \ac{wf}, a solar \ac{pv} plant, a \ac{csp} with \ac{ts}, and a \ac{fd}. Forecast data are modeled using bounds from historical records: solar \ac{pv} and \ac{csp} from CIEMAT Spain~\cite{web:ciemat_spain}, and the \ac{wf} from Iberdrola Spain~\cite{web:iberdrola_spain}. To avoid overly conservative solutions, bounds are set between the 20th and 80th percentiles. Forecast ranges for production and consumption are shown in Figure~\ref{fig:Data_Production_Robust}. \ac{fd} forecasts are based on three demand profiles from~\cite{ortega2022modeling}, each with a 10–30\% flexibility margin allowing demand to deviate above or below nominal levels. Electricity price forecasts for the \ac{dam}, \ac{srm}, and \acs{idm} rely on historical data from~\cite{REE2025} and are illustrated in Figure~\ref{fig:Data_Price_Robust}. {\color{black}Note that forecast bounds for \acp{idm} prices tighten as bidding time approaches delivery, reflecting improved forecast accuracy. The updated forecast bounds for \ac{ndrs} units in the \acp{idm} are omitted here to avoid cluttering the figures.} Technical specifications of all units are listed in Table~\ref{table:Data_Det}~\cite{ortega2022modeling}. 
Table~\ref{table:Data_Budget_Robust} presents the predefined budgets associated with the different uncertain parameters used in the case studies. {\color{black}Based on these budgets, we define three uncertainty-handling strategies for the \ac{rvpp} operator: optimistic, balanced, and pessimistic.} Since solar \ac{pv} generation and thermal output of the \ac{csp} are zero at night, their budgets are assigned smaller values. The budget for the \ac{idm}\#3 session, which spans fewer than 24 hours, is also proportionally reduced. This ensures a consistent share of uncertain hours across the simulation horizon for all parameters.

{\color{black}
Three case studies assess the impact of different components and multi-market participation on \ac{rvpp} performance:}

\begin{itemize}
{\color{black}
\item \textbf{Case 1:} Analyze simultaneous \ac{dam}+\ac{srm} operation, representative of most European market designs, and the optimal operation of \ac{rvpp} units under optimistic, balanced, and pessimistic uncertainty-handling strategies.

\item \textbf{Case 2:} Examine the optimal multi-market trading strategy of the \ac{rvpp} across sequentially cleared \ac{dam}, \ac{srm}, and \acp{idm}, typical of the Spanish market structure (Figure~\ref{fig:Scheme_RVPP}), under the three uncertainty-handling strategies.

\item \textbf{Case 3:} Assess the value of flexibility resources for \ac{rvpp} profitability using the \ac{mc} method in the \ac{dam}+\ac{srm} under the three uncertainty-handling strategies.
}
\end{itemize}

\begin{table}[t!]
  \centering
    \footnotesize
  \caption{RVPP units data.}
    \label{table:Data_Det}
  \footnotesize
  \setlength{\tabcolsep}{1.5pt}
  \renewcommand{\arraystretch}{1}
  \begin{threeparttable}
  \begin{tabular}{lccccccccccccccc}
    \toprule

    \multicolumn{1}{c}{\textbf{Parameter}}  
    && \multicolumn{1}{c}{\textbf{PV}}
    && \multicolumn{1}{c}{\textbf{WF}} 
    && \multicolumn{1}{c}{\textbf{Hydro}}
    && \multicolumn{1}{c}{\textbf{Biomass}}
    && \multicolumn{1}{c}{\textbf{CSP}}
    && \multicolumn{1}{c}{\textbf{TS}}
    && \multicolumn{1}{c}{\textbf{FD}}
    \\

    \cmidrule{1-1} \cmidrule{3-3} \cmidrule{5-5} \cmidrule{7-7} \cmidrule{9-9} \cmidrule{11-11} \cmidrule{13-13}  \cmidrule{15-15}

   \multirow{1}{*}{\text{$\bar P_{u}$/$\ubar P_{u}$ [MW]}}    && 50/0 
   && 50/0 && 50/10 && 10/2 && 55/5 && - && 180/0 \\ [0.2em]

     \multirow{1}{*}{\text{$R_{u}^{\uparrow}$/$R_{u}^{\downarrow}$ [MW/min]}}    && 3 && 2 && 10 && 2 && 10 && - && 5  \\ [0.2em]


        \multirow{1}{*}{\text{$UT_u$/$DT_u$ [hour]}}    && - && - && 1/0 && 3/3 && 3/2 && - &&  - \\ [0.2em]



        \multirow{1}{*}{\text{$\bar P_{\theta}/\ubar P_{\theta}$ [MWh]}}    && - && - && - && - && - && 140/0 && - \\ [0.2em]



        \multirow{1}{*}{\text{$\eta_\theta$ [\%]}}    && - && - && - && - && 95 && 90 && - \\ [0.2em]

        \multirow{1}{*}{\text{$C_{u}$ [€/MWh]}}    && 10 && 15 && 12.5 && 60 && 25 && - && -  \\ [0.2em]

\bottomrule
  \end{tabular}
\end{threeparttable}
\vspace{-2.5mm}
\end{table}

\begin{table}[t!]
  \centering
  \caption{Budgets for different uncertain parameters.}
  \footnotesize
    \setlength{\tabcolsep}{1.5pt} 
  \renewcommand{\arraystretch}{.8} 
  \begin{threeparttable}
  \begin{tabular}{llccc}
    \toprule
    \textbf{Uncertainty type} & \textbf{Parameter} & \textbf{Optimistic} & \textbf{Balanced} & \textbf{Pessimistic} \\
    \midrule
    \multirow{3}{*}{Market price} 
        & \ac{dam}/\ac{srm}          & 4 & 8 & 12 \\
        & \ac{idm}\#1–\ac{idm}\#2    & 4 & 8 & 12 \\
        & \ac{idm}\#3                & 2 & 4 & 6 \\
    \midrule
    \multirow{3}{*}{Renewable production}
        & \ac{wf}                     & 4 & 8 & 12 \\
        & PV                          & 2 & 4 & 6 \\
        & \ac{csp} thermal           & 2 & 4 & 6 \\
    \midrule
    \text{Load consumption} & \ac{fd}   & 4 & 8 & 12 \\
    \bottomrule
  \end{tabular}
  \end{threeparttable}
  \label{table:Data_Budget_Robust}
  \vspace{-1mm}
\end{table}



\subsection{Case 1}\label{subsec:Case1}

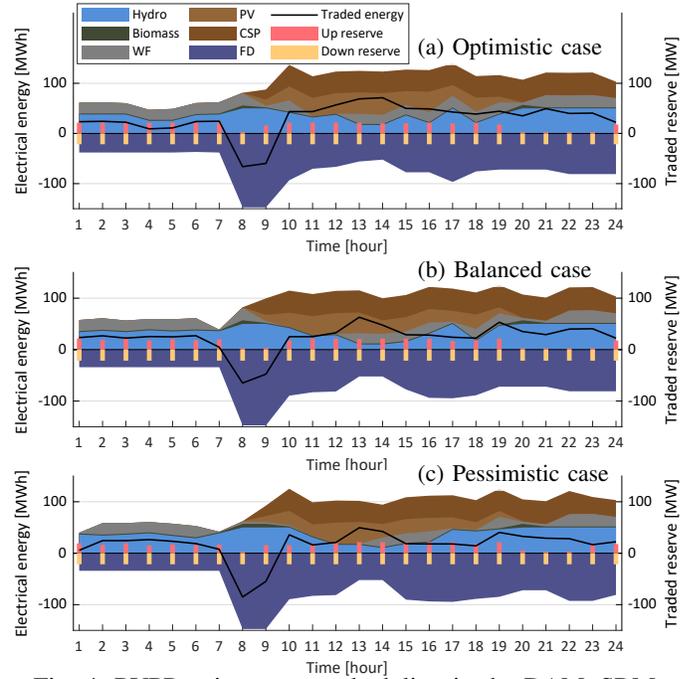
\begin{figure}[t!]
    \centering
\vspace{-.6em}
    \begin{tikzpicture}
      \node[inner sep=0] (imgA)
        {\includegraphics[width=\linewidth]{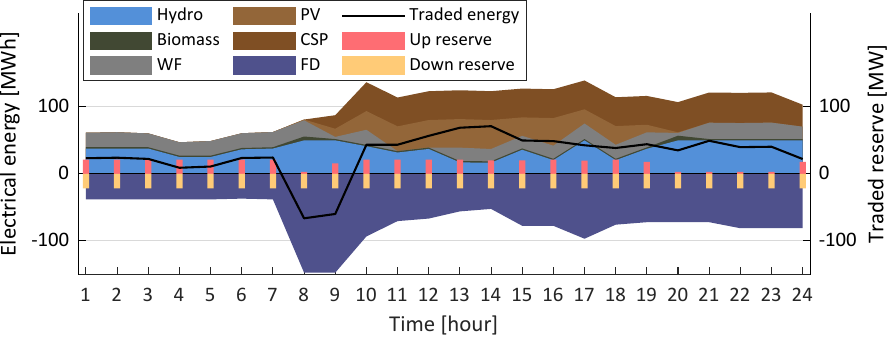}};
      \node[panel label, xshift=.55cm, yshift=-.1cm] at (imgA.north) {(a) Optimistic case};
    \end{tikzpicture}
 \vspace{-2mm} 
    \begin{tikzpicture}
      \node[inner sep=0] (imgB)
        {\includegraphics[width=\linewidth]{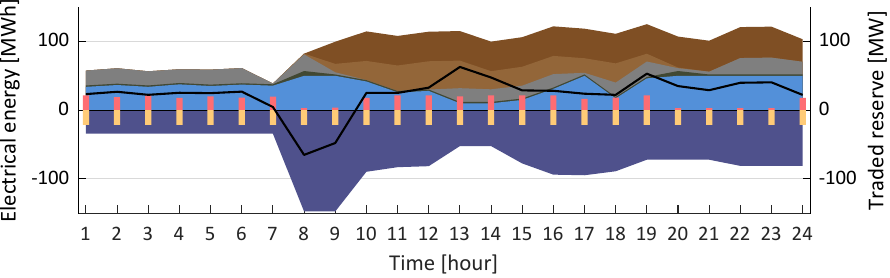}};
      \node[panel label, xshift=.55cm, yshift=.45cm] at (imgB.north) {(b) Balanced case};
    \end{tikzpicture}
 \vspace{-1.5mm} 
    \begin{tikzpicture}
      \node[inner sep=0] (imgC)
        {\includegraphics[width=\linewidth]{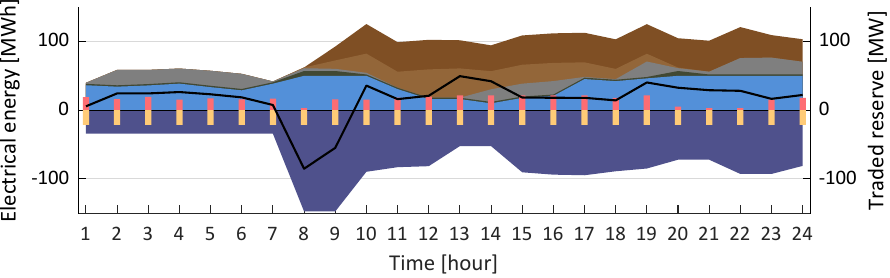}};
      \node[panel label, xshift=.55cm, yshift=.45cm] at (imgC.north) {(c) Pessimistic case};
    \end{tikzpicture}

    \caption{RVPP units energy scheduling in the DAM+SRM. 
    }
    \label{fig:Units_energy_Robust_All}
    \vspace{-1.5em}
\end{figure}

{\color{black}Figure~\ref{fig:Units_energy_Robust_All} shows the energy generation and consumption schedules of the \ac{rvpp} units, as well as the total reserve of the \ac{rvpp} in the \ac{dam}+\ac{srm}, under different strategic approaches adopted by the \ac{rvpp} operator. The results show that when moving from optimistic to more conservative (balanced and pessimistic) strategies, the utilization of renewable units—particularly the \ac{wf} and solar \ac{pv}—generally decreases during several hours. This occurs because the operator adopts a more cautious scheduling approach to hedge against forecast uncertainty, reducing reliance on variable generation and increasing the contribution from more controllable units.} 
For instance, in the optimistic case, energy fluctuations of the \ac{wf} occur in hours 9, 11, 12, and 20. In contrast, in the balanced and pessimistic cases, these fluctuations occur during hours 7, 9–12, 17, 20, and 21, and during hours 1, 7–13, 17, 18, 20, and 21, respectively. Although uncertainty in the thermal input energy of the \ac{csp} is considered, the \ac{csp} can effectively mitigate these fluctuations with the support of its \ac{ts}, resulting in only marginal impacts on its final electrical energy output. Additionally, the hydro plant plays a crucial role in compensating for energy shortages from \acs{ndrs}. For example, in hour 18, the \ac{wf} generates energy in the optimistic and balanced cases but not in the pessimistic case. This shortage in the pessimistic case is offset by higher energy generation from the hydro plant at that hour.

\begin{figure}[t!]
    \centering
\vspace{-1mm}
    \begin{tikzpicture}
      \node[inner sep=0] (imgA)
        {\includegraphics[width=0.48\textwidth]{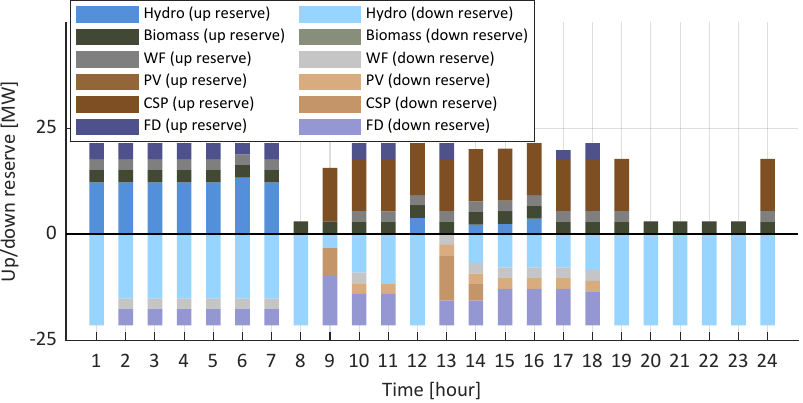}};
      \node[panel label, xshift=1.2cm, yshift=-.6cm] at (imgA.north) {(a) Optimistic case};
    \end{tikzpicture}
    \hfill
    \vspace{-1.5mm} 
    \begin{tikzpicture}
      \node[inner sep=0] (imgB)
        {\includegraphics[width=0.48\textwidth]{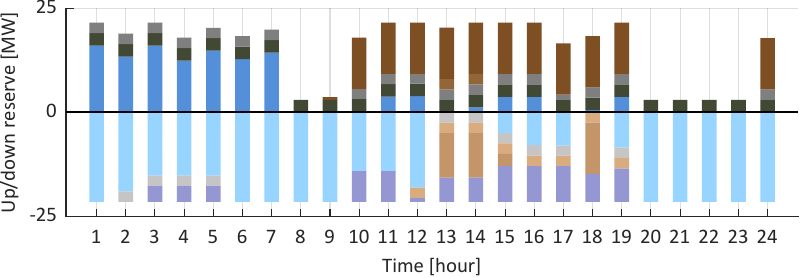}};
      \node[panel label, xshift=1.2cm, yshift=.45cm] at (imgB.north) {(b) Balanced case};
    \end{tikzpicture}

    \vspace{-1.2mm} 

    \begin{tikzpicture}
      \node[inner sep=0] (imgC)
        {\includegraphics[width=0.48\textwidth]{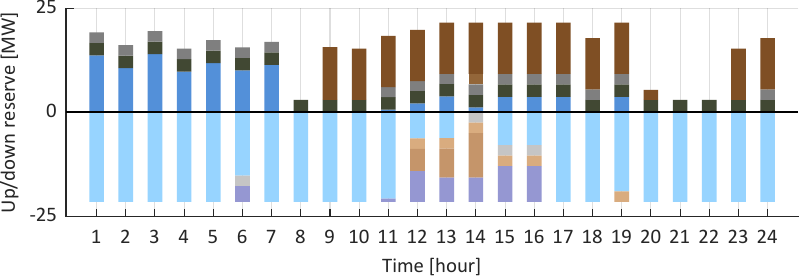}};
      \node[panel label, xshift=1.2cm, yshift=.45cm] at (imgC.north) {(c) Pessimistic case};
    \end{tikzpicture}

    \caption{RVPP units reserve scheduling in the DAM+SRM. 
    }
    \label{fig:Units_reserve_Robust_All}
    \vspace{-1.5em}
\end{figure}
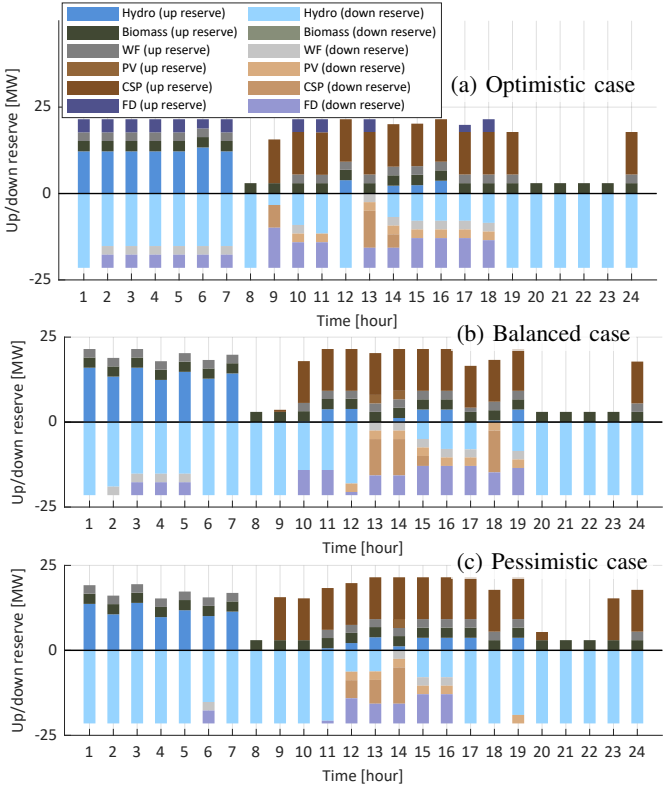

{\color{black}Figure~\ref{fig:Units_reserve_Robust_All} details the reserve provided by \ac{rvpp} units participating in the \ac{dam}+\ac{srm} under different strategic approaches of the \ac{rvpp} operator.} The results show that adopting more conservative strategies leads to a reduction in the total upward reserve provided by \ac{rvpp} units. For example, in the pessimistic case, the provision of upward reserve decreases during hours 1–7, 10–12, and 18 compared to the optimistic case. These reductions occur because greater uncertainty negatively impacts the production of \ac{ndrs} and the consumption of demand, requiring \ac{drs} to compensate for energy shortages and thereby reducing their availability for reserve provision. In all cases, the hydro plant and \ac{csp} provide the largest share of upward reserve due to their inherent flexibility. Specifically, the hydro plant contributes significantly during hours 1–7, while the \ac{csp} supplies reserve efficiently between hours 9–19 when it is available. Notably, although the \ac{fd} provides upward reserve in the optimistic case, it does not do so in the balanced and pessimistic cases, as its flexibility is primarily allocated to compensate for the energy reduction of \ac{ndrs}.

\subsection{Case 2}\label{subsec:Case2}

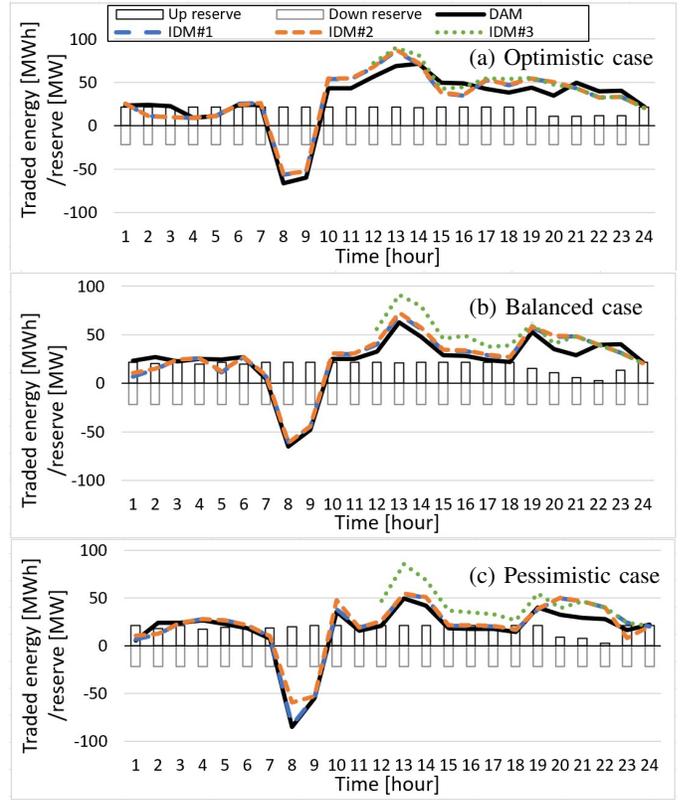
\begin{figure}[t!]
    \centering

    \begin{tikzpicture}
      \node[inner sep=0] (imgA)
        {\includegraphics[width=\linewidth]{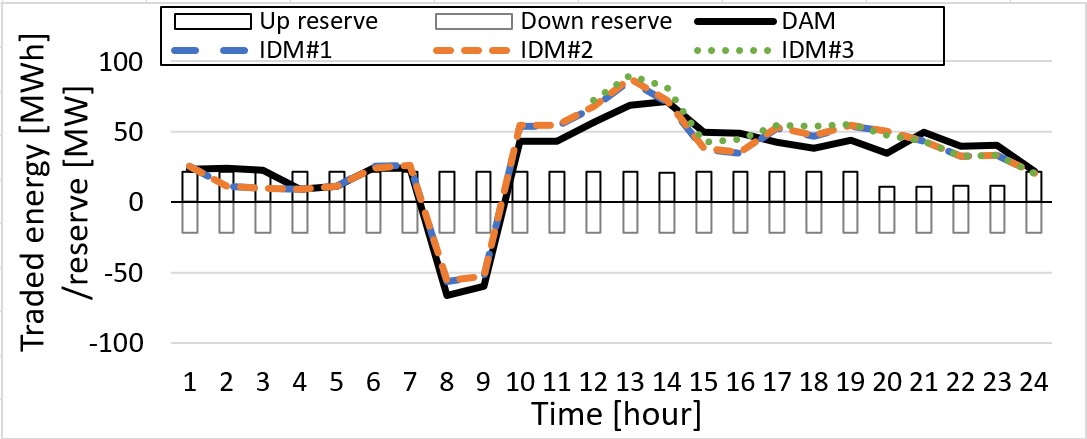}};
      \node[panel label, xshift=1.3cm, yshift=-.25cm] at (imgA.north) {(a) Optimistic case};
    \end{tikzpicture}


    \begin{tikzpicture}
      \node[inner sep=0] (imgB)
        {\includegraphics[width=\linewidth]{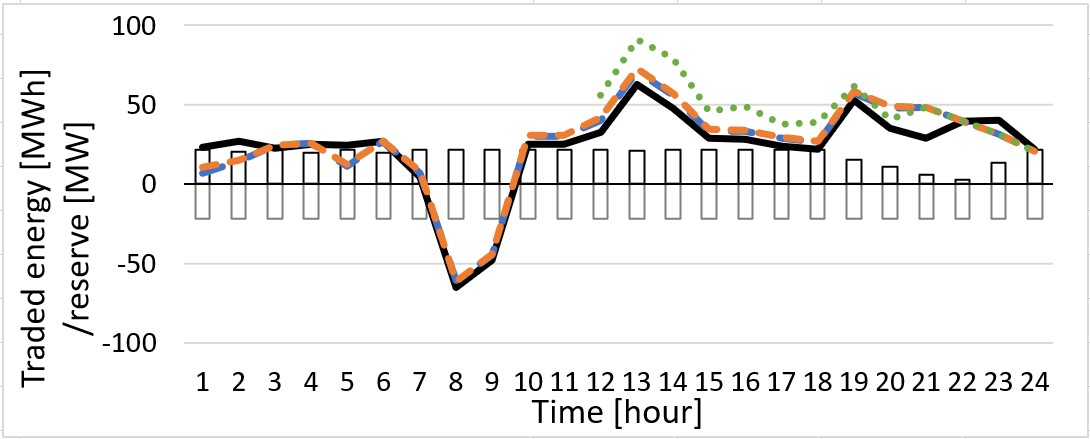}};
     \node[panel label, xshift=1.3cm] at (imgB.north) {(b) Balanced case};
    \end{tikzpicture}


    \begin{tikzpicture}
      \node[inner sep=0] (imgC)
        {\includegraphics[width=\linewidth]{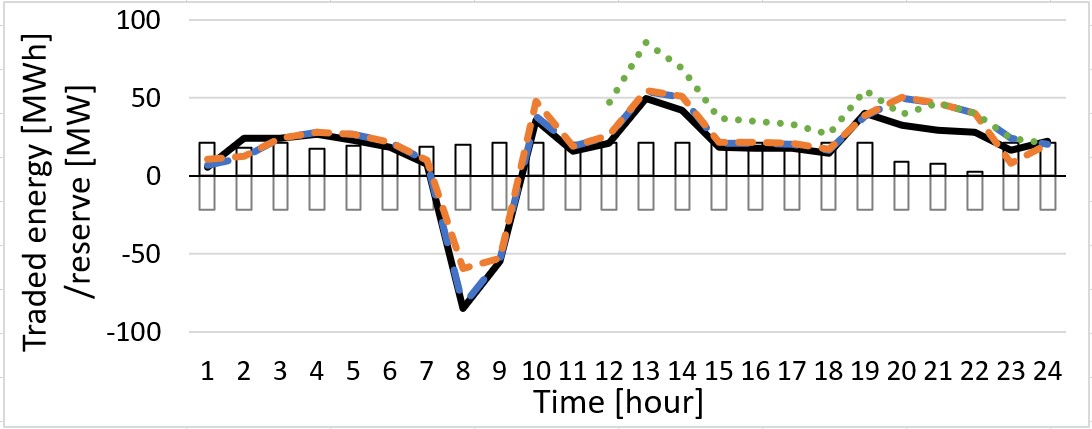}};
     \node[panel label, xshift=1.3cm] at (imgC.north) {(c) Pessimistic case};
    \end{tikzpicture}

    \caption{RVPP trading approaches across multi-market sessions. 
    }
    \label{fig:Traded_energy_Robust_All}
    \vspace{-1.5em}
\end{figure}

Figure~\ref{fig:Traded_energy_Robust_All} shows the traded energy and reserve of the \ac{rvpp} across different electricity market sessions under various strategic approaches of the \ac{rvpp} operator, with the values in each session reflecting cumulative trades that include all previous sessions. The \ac{rvpp} is primarily an energy seller in the optimistic, balanced, and pessimistic cases, except during hours 8–9. In the optimistic case, where uncertain parameters are allowed to deviate to their worst-case values in only a limited number of hours, changes in total traded energy are observed up to \ac{idm}\#3, with the most significant changes occurring in \ac{idm}\#1. In \ac{idm}\#1, the \ac{rvpp} acts as an energy seller during hours 8–13 and 17–20, and as an energy buyer during hours 2–3, 15–16, and 21–24. This leads to an increase in total energy sold during hours 10–13 and 17–20, and a decrease in both the energy sold during hours 2–3, 15–16, and 21–24, and the energy bought during hours 8–9, compared to the \ac{dam} session. In the balanced case, uncertain parameters are allowed to deviate in a greater number of hours. As a result, the energy sold by the \ac{rvpp} in the \ac{dam} decreases compared to the optimistic case, particularly in hours 7, 10–18, and 21. Additionally, the up reserve provided by the \ac{rvpp} is reduced in hours 19–22 compared to the optimistic case. Moreover, changes in the traded energy of the \ac{rvpp} are observed in all \ac{idm} sessions, and these changes are of higher magnitude than in the optimistic case. This is because, with increased uncertainty, greater fluctuations in the production of \acs{ndrs} are considered, making the role of the \ac{idm} in adjusting the \ac{rvpp}'s energy more critical. In the pessimistic case, higher uncertainty budgets are applied compared to the previous optimistic and balanced cases. This leads to further reductions in the energy sold and reserves provided across more hours, along with an increase in energy purchased during hours 8–9 to supply internal demand. Additionally, more substantial changes in the traded energy of the \ac{rvpp} are observed across the \acp{idm} compared to the earlier strategies.

Table~\ref{table:Economic_Robust_All} presents the economic results across different multi-market sessions under various uncertainty strategies. The main income of the \ac{rvpp} is derived from energy and reserve provision in the \ac{dam} and \ac{srm}, while the \acp{idm} primarily serve for energy adjustments. Under more conservative strategies, \ac{rvpp} revenue in the \ac{idm} sessions increases compared to the optimistic case. For example, the revenue change in \ac{idm}\#1 between the optimistic and the balanced and pessimistic strategies is 13.4\% and 85.8\%, respectively. These results indicate that when more uncertainty is considered, the role of the \acp{idm} in energy adjustment becomes more significant. The total profit across all market sessions, however, is lower in the conservative cases. This is because under a conservative strategy, the \ac{rvpp} submits lower bids in the energy and reserve markets, which reduces total revenue, while the costs associated with electricity price uncertainty increase. For instance, the profit of the \ac{rvpp} in the balanced and pessimistic cases decreases by 70.4\% and 128.6\%, respectively. The corresponding reductions in total revenue are 7.2\% and 12.1\%, whereas costs rise by 7.1\% and 14.4\%.

\begin{table}[t!]
  \centering
  \footnotesize
 \setlength{\tabcolsep}{5pt}
  \caption{RVPP economic results in multi-market sessions. 
  }
    \renewcommand{\arraystretch}{.9} 
  \begin{threeparttable}
    \begin{tabular}{lccc}
      \toprule
      \textbf{Metric} 
      & \textbf{Optimistic} 
      & \textbf{Balanced} 
      & \textbf{Pessimistic} \\
      
      \specialrule{1pt}{1pt}{1pt}

      DAM revenue [k€] & 32.48 & 25.61 & 17.60 \\
      SRM revenue [k€] & 25.76 & 25.25 & 25.12 \\
      IDM\#1 revenue [k€] & 2.46 & 2.79 & 4.57 \\
      IDM\#2 revenue [k€] & 0.37 & 0.89 & 1.43 \\
      IDM\#3 revenue [k€] & 0.44 & 2.56 & 5.36 \\
      Cost [k€] & 50.13 & 53.73 & 57.34 \\
      Profit [k€] & 11.38 & 3.37 & -3.26 \\
      
      \bottomrule
    \end{tabular}
  \end{threeparttable}
  \label{table:Economic_Robust_All}
  \vspace{-2em}
\end{table}

\subsection{Case 3}\label{subsec:Case3}

{\color{black}Table~\ref{table:Economic_Robust_WO_Unit} presents the profit for different \ac{rvpp} configurations in the \ac{dam}+\ac{srm}, where individual technologies are excluded. The total profit (or cost) of the \ac{rvpp} when all units are included is 8.75 k€, –1.67 k€, and –10.67 k€ under the optimistic, balanced, and pessimistic strategies, respectively.} In general, technologies with larger capacity shares, lower operating costs, and dispatchable characteristics have the greatest impact on \ac{rvpp} profitability. Excluding all \ac{drs} (hydro and biomass plants) reduces total profit by {\color{black}40.73 k€ (=\,8.75 $-$ (-31.98))}, 39.22 k€, and 38.55 k€ in the optimistic, balanced, and pessimistic cases, respectively. By contrast, removing the \ac{fd} (treating demand as inflexible) decreases profit by only {\color{black}7.42 k€ (=\,8.75 $-$ 1.33)}, 7.78 k€, and 7.44 k€, since just 10\% of demand is flexible. Excluding \ac{ndrs} (solar \ac{pv} and \ac{wf}) reduces profit by 30.44 k€, 24.17 k€, and 19.31 k€. Because the \ac{rvpp} includes a significant share of \ac{ndrs} (50~MW each of \ac{wf} and solar \ac{pv}) and these units have low operating costs, their exclusion leads to notable profit differences compared to \ac{drs}. However, the contribution of \ac{ndrs} is more volatile under conservative strategies due to their stochastic production. To enable fair comparison, the normalized contribution of each technology ($\rho_u$) is reported. The results indicate that \ac{drs} have the highest normalized contribution across all strategies, with $\rho_u$ reduced by only 4.4\% and 5.9\% in the balanced and pessimistic cases relative to the optimistic case. This confirms that excluding \ac{drs} strongly affects energy and reserve provision. The \ac{fd} makes the second-largest normalized contribution after \ac{drs}, as it provides cost-effective balancing by shifting demand. The \ac{csp} has a smaller normalized contribution than \ac{drs}, since its thermal input is subject to uncertainty, unlike \ac{drs}. For example, in the optimistic, balanced, and pessimistic cases, $\rho_u$ for the configuration without \ac{csp} is 44.1\%, 46.1\%, and 50.0\% lower, respectively, than for the configuration without \ac{drs}. Nevertheless, \ac{csp} makes a higher normalized contribution than \ac{ndrs}, as its \ac{ts} allows it to effectively manage uncertainty and enhance profitability. Notably, the $\rho_u$ of the configuration without \ac{ndrs} decreases by 20.0\% and 36.6\% in the balanced and pessimistic strategies, respectively, compared to the optimistic case, highlighting the strong effect of uncertainty on the normalized contribution of \ac{ndrs}.

{\color{black}The results in Table~\ref{table:Economic_Robust_WO_Unit} also show the profit allocation for each technology ($\Pi_u^{\mathrm{alloc}}$), and the profit each technology would earn if participating in the market individually ($\Pi_u^{\mathrm{solo}}$). For the \ac{drs}, the proposed model increases profits by 21.4\%, 36.0\%, and 52.9\% in the optimistic, balanced, and pessimistic cases, respectively. Consequently, the \ac{drs} achieve a stable profit between 43.34 k€ and 44.97 k€ across all uncertainty-handling strategies, which is reasonable given the flexibility they provide to the \ac{rvpp}. The corresponding profit increases for the \ac{ndrs} are 23.0\%, 36.0\%, and 55.5\%, respectively. However, the allocated profit for the \ac{ndrs}, ranging between 21.83 k€ and 30.56 k€, is more volatile than that of the \ac{drs} across different uncertainty-handling strategies due to their stochastic production characteristics.}

\begin{table}[t!]
  \centering
  \footnotesize
 \setlength{\tabcolsep}{5pt}
  \caption{Profit allocation results for different technologies. 
  }
    \renewcommand{\arraystretch}{.95} 
  \begin{threeparttable}
    \begin{tabular}{llcccc}
      \toprule
      \textbf{Strategy} & \textbf{Technology} 
      & \text{$\rho_u$}
      & \text{$\Pi^{\mathrm{RVPP}\setminus u}$}
      & \text{$\Pi_u^{\mathrm{solo}}$}
      & \text{$\Pi_u^{\mathrm{alloc}}$}  \\

    \multicolumn{1}{c}{\textbf{}} & {\textbf{}} & \text{[k€/MW]} & \text{[k€]} & \text{[k€]}  & \text{[k€]} 
    \\
      
      \specialrule{1pt}{1pt}{1pt}

      \multirow{4}{*}{Optimistic} 
      & D-RES     & 0.68    & -31.98  & 35.70  & 43.34  \\ 
      & CSP       & 0.38  & -12.32  & 17.19  & 21.14 \\ 
      & ND-RES    & 0.30   & -21.69  & 24.85  & 30.56   \\
      & FD        & 0.41   & 1.33  & -87.69 & -86.30   \\
      \specialrule{.3pt}{.3pt}{.3pt}

            \multirow{4}{*}{Balanced} 
      & D-RES   &  0.65   & -40.89  & 32.39  & 44.05  \\ 
      & CSP     &  0.35  & -21.00 & 13.33 &  19.08 \\ 
      & ND-RES  & 0.24   & -25.84  & 18.42 & 25.60  \\
      & FD      & 0.43    & -9.45  & -92.71 & -90.40   \\ 
     \specialrule{.3pt}{.3pt}{.3pt}
      \multirow{4}{*}{Pessimistic} 
      & D-RES    & 0.64  & -49.22  & 29.41 & 44.97  \\ 
      & CSP      & 0.32  & -28.43  & 10.62 & 17.79   \\ 
      & ND-RES   & 0.19  & -29.98  & 14.04  & 21.83  \\ 
      & FD       & 0.41  & -18.11  & -98.26 & -95.26  \\

      \bottomrule
    \end{tabular}
    \vspace{-1mm}
  \end{threeparttable}
  \label{table:Economic_Robust_WO_Unit}
\end{table}

 \begin{table}[t!]
  \centering
  \footnotesize
 \setlength{\tabcolsep}{2pt}
  \caption{Economic results 
  for varying demand flexibility. 
  }
    \renewcommand{\arraystretch}{.91} 
  \begin{threeparttable}
    \begin{tabular}{lccccc}
      \toprule
      \textbf{Strategy} & \textbf{Demand} 
      & \text{DAM}
      & \text{SRM}
      & \text{Cost}
      & \text{Profit} \\

    \multicolumn{1}{c}{\textbf{}} & {\textbf{flexibility [\%]}} & \text{revenue [k€]} & \text{revenue [k€]} & \text{[k€]}  & \text{[k€]} 
    \\
      
      \specialrule{1pt}{1pt}{1pt}

      \multirow{4}{*}{Optimistic} 
      & 0     & 24.67   & 23.71  & 47.05 & 1.33  \\
      & 10    & 32.48   & 23.79  & 47.53 & 8.75 \\ 
      & 20    & 38.69   & 24.99  & 50.47 & 13.21 \\ 
      & 30    & 36.63   & 26.28  & 47.23  & 15.69 \\ 
     \specialrule{.3pt}{.3pt}{.3pt}

      \multirow{4}{*}{Balanced} 
      & 0     & 17.45    & 22.73  & 49.63 & -9.45  \\ 
      & 10    & 25.61    & 22.75  & 50.03 & -1.67  \\ 
      & 20    & 31.69   & 24.31  & 52.37  & 3.63 \\
      & 30    & 31.26   & 25.14  & 49.98  & 6.42 \\ 
     \specialrule{.3pt}{.3pt}{.3pt}

      \multirow{4}{*}{Pessimistic} 
      & 0     & 9.45    & 22.81  & 50.37 & -18.11  \\ 
      & 10    & 17.60    & 22.93  & 51.20 & -10.67  \\
      & 20    & 24.91   & 23.50  & 52.94  & -4.53 \\ 
      & 30    & 24.96   & 24.07  & 50.44  & -1.41 \\ 
      \bottomrule
    \end{tabular}
  \end{threeparttable}
  \label{table:Economic_Robust_FD}
  \vspace{-1.5em}
\end{table}


{\color{black}In the previous case studies, only 10\% demand flexibility was considered, and although its direct impact on the \ac{rvpp} profit was modest, the \ac{fd} showed a substantial contribution to profit improvement. Therefore, additional simulations are performed to assess how higher flexibility levels influence \ac{rvpp} performance under different uncertainty-handling strategies.} In this regard, Table~\ref{table:Economic_Robust_FD} presents the \ac{rvpp} profit in the \ac{dam}+\ac{srm} for different levels of demand flexibility. The results show that increasing demand flexibility generally leads to higher \ac{rvpp} profit, with the effect being more pronounced under conservative strategies. For instance, when flexibility increases from 0\% to 30\%, profit rises by 14.36 k€, 15.87 k€, and 16.70 k€ in the optimistic, balanced, and pessimistic cases, respectively. These findings highlight the importance of \ac{fd} in enhancing the energy and reserve provision of the \ac{rvpp}, particularly when uncertainties strongly affect its performance.


\section{Conclusions}
\label{sec:Conclusions}

This paper analyzes the impact of different flexible resources, including \ac{drs}, \ac{csp}, and \ac{fd}, on the multi-market participation and profitability of an \acs{rvpp} with a high share of \ac{ndrs}. The analysis incorporates uncertainties in energy and reserve market prices, renewable generation, and demand consumption using a two-stage robust approach. A marginal contribution method—accounting for each unit’s actual contribution to energy and reserve provision as well as the final profit of the \ac{rvpp}—is applied to allocate the additional profit of the \ac{rvpp} (compared to individual unit participation) among its units. Simulation results show how the \ac{rvpp} operator schedules units and submits energy and reserve bids under optimistic, balanced, and pessimistic strategies. More conservative strategies reduce energy sold and increase energy purchased, providing robustness against worst-case scenarios. Furthermore, in the balanced and pessimistic cases, changes in the traded energy of the \ac{rvpp} across \acs{idm} sessions are greater than in the optimistic case, highlighting a stronger need for adjustments in the \acs{idm} and reinforcing their importance. Additionally, the hydro plant and \ac{csp} compensate for energy shortages from \ac{ndrs} under more conservative strategies, which in turn reduces their reserve provision. The results also show that \ac{drs} and \ac{fd} make the highest normalized contributions to \ac{rvpp} profitability, and these contributions remain relatively stable across different uncertainty-handling strategies. By contrast, \ac{ndrs} make smaller normalized contributions with greater volatility under conservative strategies, since their output is strongly affected by generation stochasticity. The \ac{csp} provides a moderate normalized contribution: although its production is influenced by thermal uncertainty, its thermal storage effectively mitigates this impact.



%





\bibliographystyle{IEEEtran}
\bibliography{refs.bib}


\end{document}